%% file: main.tex
\documentclass[journal]{IEEEtran}

\usepackage{amsmath,graphicx}
\usepackage[usenames, dvipsnames]{xcolor}

\input{setup/packages}
\input{setup/macros}

\begin{document}

\makeatletter
\newcommand{\mypm}{\mathbin{\mathpalette\@mypm\relax}}
\newcommand{\@mypm}[2]{\ooalign{%
  \raisebox{.1\height}{$#1+$}\cr
  \smash{\raisebox{-.6\height}{$#1-$}}\cr}}
\makeatother

\title{Camera Fingerprint Extraction via Spatial Domain Averaged Frames}

\author{Samet Taspinar,
        Manoranjan Mohanty,
        and Nasir Memon
\thanks{Samet Taspinar (email: st89@nyu.edu) and Manoranjan Mohanty email: manoranjan.mohanty@nyu.edu) are with Center for Cyber Security, New York University Abu Dhabi, UAE. Nasir Memon (email: nm1214@nyu.edu) is with Department of Computer Science and Engineering, New York University, New York, USA.}
}

\markboth{IEEE TRANSACTIONS ON INFORMATION FORENSICS AND SECURITY}%
{\framework}

\maketitle

\input{sections/1_abstract}

\begin{keywords}
PRNU, video forensics, camera fingerprint extraction, image forensics.
\end{keywords}

\input{sections/2_introduction}

\input{sections/3_background}

\input{sections/4_our_approach}

\input{sections/5a_exp_validation.tex}
\input{sections/5b_exp_video.tex}

\input{sections/7_conclusion}

\bibliographystyle{IEEEtran}
\bibliography{biblio.bib}

\end{document}

%% file: setup/packages.tex
\usepackage{cite}
\usepackage[pass]{geometry}
\usepackage[ruled]{algorithm2e}
\usepackage{subfig}
\usepackage{array}

\usepackage{makeidx}         
\usepackage{graphicx}        
\usepackage{multicol}        
\usepackage[bottom]{footmisc}
\usepackage{amssymb,amsmath,amsthm,amsfonts}

 \usepackage[hyphens]{url}
\usepackage{breakurl}

\usepackage{dsfont}

\usepackage{todonotes}
\usepackage{color}

\usepackage{multirow}

%% file: setup/macros.tex
\SetAlFnt{\small}
\SetAlCapFnt{\small}
\SetAlCapNameFnt{\small}
\SetAlCapHSkip{0pt}
\IncMargin{-\parindent}

\usepackage{xspace}

\mathchardef\mhyphen="2D

\newcommand{\framework}{\textsl{\mbox{2DCrypt}}\xspace}

\renewcommand{\paragraph}[1]{\medskip \noindent \textbf{#1.\ }}

%% file: sections/1_abstract.tex
\begin{abstract}
Photo Response Non-Uniformity (PRNU) based camera attribution is an effective method to determine the source camera of visual media (an image or a video). 
To apply this method, images or videos need to be obtained from a camera to create a ``camera fingerprint" which then can be compared against the PRNU of the query media whose origin is under question.  The fingerprint extraction process can be time consuming when a large number of video frames or images have to be denoised. This may need to be done when the individual images have been subjected to high compression or other geometric processing such as video stabilization. This paper investigates a simple, yet effective and efficient technique to create a camera fingerprint when so many still images need to be denoised. The technique utilizes Spatial Domain Averaged (SDA) frames. An SDA-frame is the arithmetic mean of multiple still images. When it is used for fingerprint extraction, the number of denoising operations can be significantly decreased with little or no performance loss. Experimental results show that the proposed method can work more than $50$ times faster than conventional methods while providing similar matching results.
\end{abstract}

%% file: sections/2_introduction.tex
\section{Introduction}
\label{sec:intro}

PRNU-based source camera attribution is a well-studied and successful method in media forensics for finding the source camera of an anonymous image or video \cite{lukas2006digital}. 
The method is based on the unique Photo Response Non Uniformity (PRNU) noise of a camera sensor array stemming from manufacturing imperfections. This PRNU noise can act as a camera fingerprint. The PRNU approach is often used in two scenarios: camera verification and camera identification. Camera verification aims to establish if a given query image or a video is taken by a suspect camera. This is done by correlating the noise estimated from the query image or video with the fingerprint of the camera usually is computed by taking pictures from the camera under controlled conditions. In camera identification, the potential source camera of the query image or video is determined from a large database of camera fingerprints. One can view camera identification as essentially the same as performing $n$ camera verification tasks where $n$ is the number of camera fingerprints in the database. However, when performing identification, it is assumed that the camera fingerprints are pre-computed. 

In both verification and identification, it is often the case that there is no camera available to create fingerprints under controlled conditions. Rather, camera fingerprints are estimated from a set of publicly available media assumed to be from the same camera. Such media can have a very diverse range of quality and content and often lacks metadata.


For efficient fingerprint matching in large databases, various approaches have been proposed. Fridrich et al. \cite{goljan2010managing} proposed the use of fingerprint digests in which a subset of fingerprint elements having the highest sensitivity are used instead of the entire fingerprint. Bayram {\em et al.} \cite{bayram2012efficient} introduced binarization where each fingerprint element is represented by a single bit. Valsesia {\em et al.} \cite{valsesia:ranproj} proposed the idea of applying random projections to reduce fingerprint dimension. Bayram et. al. \cite{bayram:composite} introduced group testing via composite fingerprint that focuses on decreasing the number of correlations rather than decreasing the size (storage) of a fingerprint. Recently, Taspinar et al. \cite{taspinar2017fast} proposed a hybrid approach that utilizes both decreasing the size of a fingerprint and the number of correlations.
All these methods were designed and tested for images, however they can also be used for videos.

Although the image-centric PRNU-based method can be extended to video \cite{chuang2011exploring, taspinar2016source, Chen:2013:Video}, source camera attribution with video presents a number of new challenges. First, a video frame is much more compressed than a typical image. Therefore, the PRNU signal extracted from a video frame is of significantly lower quality than one obtained from an image. As a result, a larger number of video frames are required to compute the fingerprint. In fact, Chuang et. al. \cite{chuang2011exploring} found that it is best to use all the frames instead of using only the I- or P-frames to compute a fingerprint. Using a large number of frames can introduce significant computation overhead. For example, computing a fingerprint from $60$ I-frames of a one-minute HD video requires one to two minutes, whereas $30$ to $40$ minutes is required if all frames are used. 

In the case of camera identification, the amount of computation can be prohibitive in practical scenarios. For example, for computing fingerprints from a thousand one-minute Full HD videos (using all $1800$ frames) using a PC may take more than $20$ days. Clearly, with billions of media objects uploaded every day on the Internet, large scale camera source identification becomes quickly infeasible. 
Although camera fingerprints stored in a database may have to be computed just once by a system, computing a fingerprint estimate at run-time from a query video can be prohibitive when faced with a reasonable number of query videos presented to the camera identification system in a day.

Besides source camera identification, digital stabilization operations performed within modern cameras also present a significant challenge for PRNU-based source camera verification for video \cite{taspinar2016source, iuliani2017hybrid, mandelli2019facing}. Video stabilization results in sensor-pixel mis-alignments between individual frames of the video as the geometric transformations performed to compensate for camera motion and spatially align each frame are different. An accurate camera fingerprint cannot be obtained using mis-aligned frames as is done with non-stabilized video even if video quality is very high. Although there are some preliminary methods that address source camera verification for stabilized video, \cite{taspinar2016source, iuliani2017hybrid}, these methods are either limited in scope or have low performance (low true positive rate) and high computation overhead. An alternate approach to address the stabilization issue for a fairly long video (at least a couple of minutes) \cite{sri2018stabilization} is to use a large number of frames for computing the fingerprint. The idea being that with a large number of frames, there will be sufficient number of aligned pixels at each spatial location that can result in the computation of an accurate fingerprint. As discussed above, this approach however, can again introduce high computation overhead unsuitable for practical use. 

As a third example, modern devices such as smartphones capture different types of media with different resolutions. For example, most cameras don't use the full sensor resolution when capturing a video and downsize the sensor output to a lower resolution by proprietary and often unknown in-camera processing techniques. For such a challenging task PRNU based source camera matching may often fail if only I-frames are used.

This paper proposes a computationally efficient way to compute a camera fingerprint from a large number of media objects, such as individual frames of a video or a large number highly compressed images taken from a social media platform. In contrast to the two-step conventional fingerprint computation method (which first estimates PRNU noise from each frame using a denoising filter and then averages several estimated individual PRNU noise estimates to get a reliable fingerprint estimate), the proposed method uses a three step approach: frame averaging, denoising, and noise averaging. The frame averaging step gets the arithmetic mean of the frames in spatial domain, resulting in a 
\textit{Spatial Domain Averaged frame (SDA-frame)} (Figure~\ref{fig:avgframe}). 
Then, in the second step each SDA-frame is denoised, and an averaging of the estimated PRNU noise is done to arrive at the final fingerprint estimate. The goal here is to minimize the number of denoising operations (as denoising is most expensive step), and also get rid of scene dependent noise by averaging multiple frames.
Experiments with VISION dataset~\cite{shullani2017vision} and NYUAD-MMD~\cite{taspinar2019source} show that the proposed method provides significant speed up in computing accurate fingerprints. It achieves significantly higher true positive rate than a fingerprint computed by I-frames only and much lower computation cost than a fingerprint obtained from all available frames while yielding similar performance.

The rest of the paper has been organized as follows. Section~\ref{sec:background} summarizes the PRNU-based method and provides an overview of how digital video stabilization works. Section~\ref{sec:approach} explains the proposed fingerprint extraction method using SDA-frames as well as an analysis comparing it with the conventional approach. The insights obtained from the analysis are experimentally validated in Section~\ref{sec:exp:valid}. Section~\ref{sec:exp:application} examines applications for which SDA-frames based technique can be used and reports the improvement that can be achieved using an SDA-based method for those cases. Section~\ref{sec:conclusion} section provides a discussion on future work and concludes the paper.

%% file: sections/3_background.tex
\section{Background and Related Work}
\label{sec:background}

In this section, we provide a brief review of PRNU-based source camera attribution and video stabilization.

\subsection{PRNU-based Source Camera Attribution}
\label{sec:background:PRNU}
PRNU-based camera attribution is established on the fact that the output of the camera sensor, $I$, can be modeled as 
\begin{equation}
 I = I^{(0)} + I^{(0)}K + \psi
\label{Eq:im_noise}
\end{equation}
where $I^{(0)}$ is the noise-free still image, $K$ is the PRNU noise, and $\psi$ is the combination of additional noise, such as readout noise, dark current, shot noise, content-related noise, and quantization noise. The multiplicative PRNU noise pattern, $K$, is unique for each camera and can be used as a camera fingerprint which enables the attribution of visual media to its source camera.
Using a denoising filter $F$ (such as a Wavelet filter) on a set of images (or video frames) of a camera,
we can estimate the camera fingerprint by first getting the
noise residual, $W_k$, (i.e., the estimated PRNU) of the $k^{th}$ image as $W_k = I_k - \hat{I}^{(0)}_k, \hat{I}^{(0)}_k = F(I_k)$, and then averaging the noise residuals of all the images.
For determining if a specific camera has taken
a given query image, we first obtain the noise residual of the query image using $F$ and then correlate the noise residual with the camera fingerprint estimate.

For images, the PRNU-based method has been well studied. Following the seminal work in~\cite{lukas2006digital}, much research has been done to improve the scheme~\cite{jessica:camera, Sutcu:2007:Improve, Li:2012:Color, Chierchia:2010:IDP, Li:2010:Enhance}, and also make camera identification effective in practical situations \cite{goljan2010managing, bayram2012efficient, taspinar2017fast, yaqub2018towards, bayram:composite}. Researchers have also studied the effectiveness of the PRNU-based method by proposing various counter forensics and anti-counter-forensics methods~\cite{Bayram:SeamCarve:2013, taspinar2016prnu} It has also shown that the PRNU method can withstand a multitude of image processing operations, such as cropping, scaling \cite{jessica:crop}, compression \cite{alles2008source, kurt:prnu}, blurring \cite{alles2008source}, and even printing and scanning \cite{ lukas:printed}.

In contrast, there has been lesser work dedicated to PRNU-based camera attribution from a video~\cite{Simone:VideoOv:2012}.
Mo Chen et al.~\cite{Chen:2007:Vid} first extended PRNU-based approach to camcorder videos. They used Normalized Cross-Correlation (NCC) to correlate fingerprints calculated from two videos, as the videos may be subject to translation shift, e.g., due to letter-boxing.
To compensate for the blockiness artifacts introduced by heavy compression (such as MPEG-x and H26-x compression), they discard the boundary pixels of a block (e.g., a JPEG block).
In~\cite{McCloskey:2008:Confidence}, McCloskey proposed a confidence weighting scheme that can improve PRNU estimation from a video by minimizing the contribution from regions of the scene that are likely to distort PRNU noise (e.g., excluding high-frequency content). Chuang et al.~\cite{chuang2011exploring} studied PRNU-based source camera identification problem with a focus on smart-phone cameras. Since smart-phones are subject to high compression, they considered only I-frames for fingerprint calculation and correlation. Chen et al.~\cite{Chen:2013:Video} proposed a method to find PRNU noise from wireless streaming videos, which are subject to blocking and blurring. In their approach, they divided a video frame into multiple blocks and did not consider the blocks having significant blocking or blurring artifacts. Chaung et al. \cite{chuang2011exploring} showed that the best possible fingerprint could be computed when all the frames are considered (instead of using only the I- or P-frames). However, to the best of our knowledge, efficient computation of fingerprint from a given video is a relatively unexplored area.

\subsection{Affine Transformation in Video Stabilization}
\label{sec:VidCap}

\begin{figure}[ht!]
\centering
 \includegraphics[height=56mm, width=85mm]{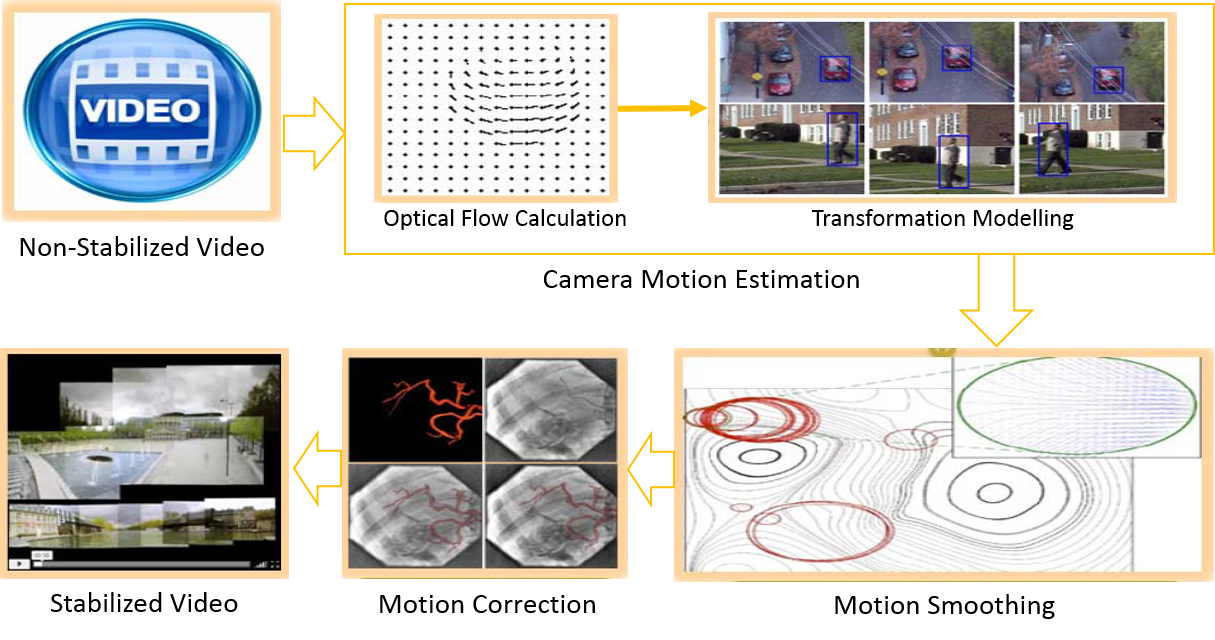}
 \caption{Video Stabilization Pipeline. This figure is a modified version of a figure that appeared in~\cite{Ejaz:2012:Video}.}
 \label{Fig:VidStab}
\end{figure}

An out-of-camera digital video stabilization process contains three major stages: camera motion estimation, motion smoothing, and motion correction (Figure~\ref{Fig:VidStab})~\cite{Matsushita:VideoStab:2006}~\cite{Ejaz:2012:Video}. In the motion estimation step, the global inter-frame motion between adjacent frames of a non-stabilized video is modeled from the optical flow vectors of the frames using an affine transformation. 
In the motion smoothing step, unintentional translations, rotations, shearing, are filtered out from the global motion vectors using a low pass filter. Finally, in the motion correction step, stabilized video is created by shifting, rotating, shearing, or zooming frames according to the parameters in the filtered motion vector. Since each video frames can use different parameters, pixels can be misaligned with the sensor array. For example, one frame can be rotated with an angle -1 degree while another by 0.5 degrees.

Digital video stabilization presents a big challenge for PRNU-based camera attribution. The frame specific affine transformations described above make the PRNU method ineffective as there is misalignment between frames. 
The brute-force methods \cite{taspinar2016prnu, iuliani2017hybrid} proposed to address the stabilization issue have had limited success and resulted in low performance. These brute-force methods try to overcome the desynchronization issue by first finding the stabilization parameters through an exhaustive search and then performing the corresponding inverse affine transformation. Such methods, therefore, have very high computation overhead.
Recently, Mandelli et al.~\cite{mandelli2019facing} improved over brute-force approaches by using a \textit{best-fit reference frame} in the parameter searching process rather than using the first frame of the given video. The \textit{best-fit reference frame} is obtained by looking for a frame that matches with the largest number of frames. Their approach also has high computation overhead.

%% file: sections/4_our_approach.tex
\section{Spatial Domain Averaging}
\label{sec:approach}
\begin{figure}[!ht]
\centering
\subfloat [1st] {\includegraphics[height=15mm]{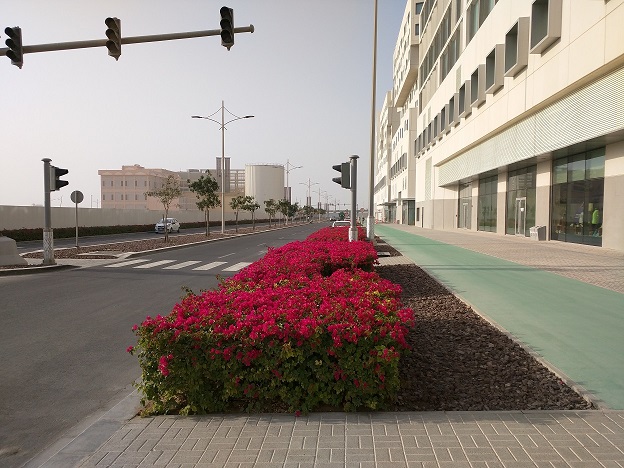}}~
\subfloat [2nd] {\includegraphics[height=15mm]{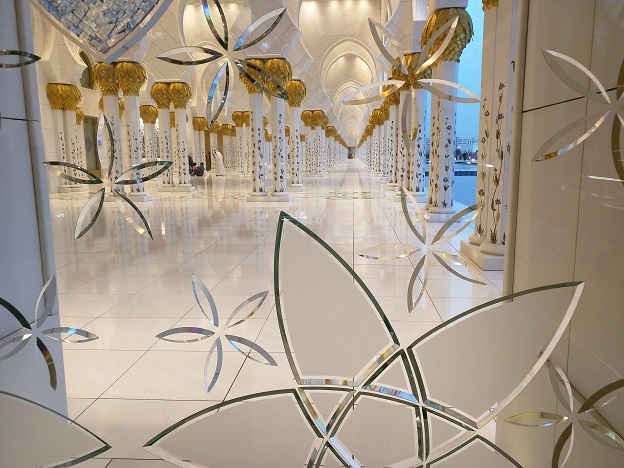}}~
\subfloat [3rd] {\includegraphics[height=15mm]{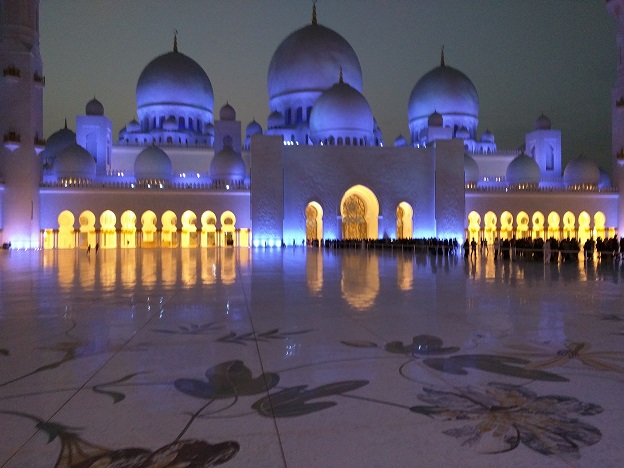}}~~
\subfloat [SDA-frame] {\includegraphics[height=15mm]{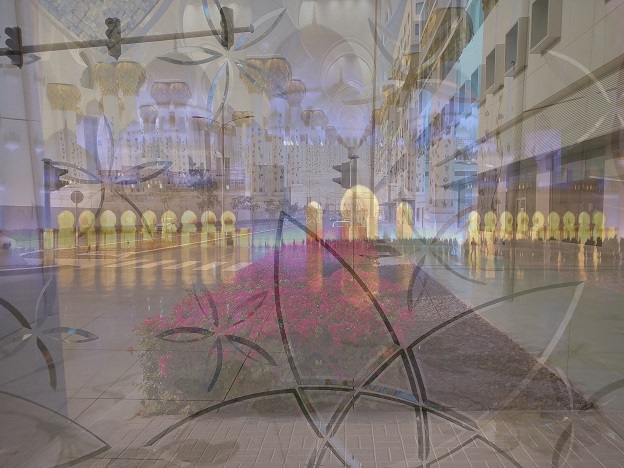}}
\caption{SDA-frame is the average of $1^{st}$, $2^{nd}$, and $3^{rd}$ frames.}
\label{fig:avgframe}
\end{figure}

As mentioned in the introduction, this paper proposes spatial domain averaging for computing camera fingerprints, which reduces the number of denoising operations when many visual objects are available. 
In the proposed method, efficient computation of a fingerprint is achieved by first creating averaged frames from a large collection, and using these averaged frames for computing the fingerprint. 
For example, given a video with $m$ frames, $g$ non-intersecting equal-sized subgroups are formed each with $d = \frac{m}{g}$ frames. A \textit{Spatial Domain Averaged frame (SDA-frame)} is created from each subgroup by getting the mean of the $d$ frames in the subgroup.
Then, in the second step, each SDA-frame is denoised, and an averaging of the estimated PRNU noise patterns is done to arrive at the final camera fingerprint estimate. In this manner, the number of frames that are denoised gets reduced by a factor of $d$. An SDA-frame obtained from three different images is shown in Figure~\ref{fig:avgframe}. 

The proposed method is inspired by the fact that although the denoising filter is designed to remove random noise from an image originating from the camera sensors (e.g., readout noise, shot noise, dark current noise etc.), as well as noise caused by processing (e.g., quantization and compression), it is not able to do a perfect job. Therefore, some scene content leaks into the extracted noise pattern. Averaging in the spatial domain acts as a preliminary filter that smoothens the image and potentially reduces the content noise that leaks into the extracted noise pattern. Of course, the effectiveness of the approach then depends on the nature of the two noise signals. Below we analyzed this fact and characterized the relationship between the noise signal arrived at by using the conventional approach and the SDA-approach. 

Further, when using the proposed approach, many questions arise. First, does frame-averaging lead to a drop in the accuracy of the fingerprint computed as compared to the conventional method, assuming the same number of images are used for both? If so, what is the trade-off between the decrease in computation and the loss in the accuracy? Can accuracy be increased by utilizing more images in the SDA method? If so, what is the optimal combination of averaging and denoising that leads to the least computation while yielding the best performance? Then, we investigated these questions, both theoretically and experimentally. We first provide a mathematical analysis using a simple framework in the two subsections below. We then validate our study in the next section by providing experimental results. The results show that spatial domain averaging strategy can indeed result in significant savings in computation while maintaining performance and in some cases, improving it.

The rest of this section provides an analysis of spatial domain averaging. To this end, we first provide an analysis of the conventional method and then analyze the SDA method.

\subsection{Conventional method}
As discussed in Section~\ref{sec:background}, in the conventional method, the camera fingerprint is estimated from $n$ images from a known camera. Each image $I$ can be modeled as $I = I^{(0)} + I^{(0)}K + \psi$, where $\psi$ is the random noise accumulated from a variety of sources
(as in (\ref{Eq:im_noise})) and $K$ is the PRNU noise.

To estimate $K$, a denoising filter, $F$, such as \cite{mihcak1999low}, BM3D\cite{dabov2009bm3d}, is used to estimate the noise free signal $I^{(0)}$. Using such a filter, we denote the noise residual as $W = I^{(0)}K + \psi + \xi$, where $\xi$ is the content noise. This noise is essentially due to sub-optimal denoising filter that is unable to completely eliminate the content from PRNU noise. 
Then, from the $n$ known image, the camera fingerprint estimate, $\hat{K}$, can be obtained using Maximum Likelihood Estimation (\textit{MLE}) as
\begin{equation}
 \hat{K} = \frac{\sum_{i=1}^{n} W_i . I_i}{\sum_{i=1}^{n} I_i^2}
\end{equation}
where $W_i$ is noise pattern extracted from $I_i$.

Note that in the estimated camera fingerprint, $\hat{K}$, $\psi$ and $\xi$ are the unwanted noise. The quality of $\hat{K}$ can be assessed from its variance $Var(\hat{K})$ \cite{Jessica:Sensor:2013}. The lower the variance is (i.e., images with smooth content), the higher the quality becomes. Assuming that $\psi$ and $\xi$ are independent White Gaussian Noise with variances $\sigma_1^2$ and $\sigma_2^2$ respectively, $Var(\hat{K})$ can be found as (using Cramer-Rao Lower Bound as shown by Fridrich et al.\cite{Jessica:Sensor:2013})
\begin{align} \label{eq:conv1}
 Var(\hat{K}) \ge \frac{\sigma_1^2 + \sigma_2^2}{\sum_{i=1}^n I_i^2}.
\end{align}
Thus a better PRNU is obtained from lower $\sigma_1^2$ and $\sigma_2^2$ (i.e., high luminance and and low textured image~\cite{Jessica:Sensor:2013}).

\subsection{Proposed SDA method}
In this subsection, we derive the variance of the estimated camera fingerprint obtained using frame averaging. We then compare this variance with that obtained by the conventional approach (in (\ref{eq:conv1})).

Suppose $I_1, I_2, \dots, I_m$ are $m$ images used to compute the camera fingerprint using SDA method. With frame averaging, these $m$ images are divided into $g = \frac{m}{d}$ disjoint sets of equal size with $d$ pictures in each set. From each set, an SDA-frame is computed. Thereafter, the process is similar to the conventional approach. Each SDA-frame is denoised, and the camera fingerprint is computed from $g$ noise residuals using MLE.

Suppose, $I_i^{SDA}$ is the SDA-frame obtained from the $i^{th}$ image set. Then 
\begin{equation}
\begin{aligned}
I_i^{SDA} &= \frac{\sum_{j=(i-1)d+1}^{id} I_j}{d} \nonumber \\
 &= \frac{\sum_{j=(i-1)d+1}^{id} (I^{(0)}_j + I^{(0)}_jK +\psi_{j}) }{d} \nonumber
\end{aligned}
\end{equation}

We can write the above equation as 
\begin{equation}
 I_i^{SDA} = I_i^{(0), SDA} + I_i^{(0), SDA}K + \psi_{i}^{SDA},
 \label{eq:avg1}
\end{equation}
where $I_i^{(0), SDA}$ is the noise free image, and $\psi_{i}^{SDA}$ is the random noise (from pre-filtering sources) in the SDA-frame. 
This noise can be written as
$$ \psi_{i}^{SDA} = \frac{\sum_{j=(i-1)d+1}^{id} \psi_{j}}{d}.
$$
Suppose $\sigma_1^2$ is the variance of $\psi$'s (which is assumed to be White Gaussian Noise). Then, the variance of $\psi_{i}^{SDA}$ turns out to be $\frac{\sigma_1^2}{d}$.

Suppose $W^{SDA}$ is the noise residual of each SDA-frame, $I^{SDA}$. Then, 
\begin{align}
W^{SDA} &= I^{SDA} - F(I^{SDA}) \nonumber \\
 &= I^{(0), SDA}K + \psi^{SDA} + \xi^{\prime}\nonumber,
\end{align} 
where $F$ is the denoising filter, and $\xi^{\prime} = I^{(0), SDA} - F(I^{SDA})$ is the content noise due to the sub-optimal nature of the denoising filter. Note that $\xi^{\prime}$ is assumed to be independent of PRNU signal $I^{(0), SDA}K$ (although $\xi^{\prime}$ contains content layover $I^{(0), SDA} - F(I^{SDA}$) as $\xi^{\prime}$ is negligible compared to $I^{SDA}_{0}K$\cite{Jessica:Sensor:2013}. 

We know that $\xi^{\prime}$ is dependent on the smoothness of the SDA-frames. If the frames contain textured content, $\xi^{\prime}$ is high. 
Assuming that SDA-frames have similar smoothness to the input frames from which they are created, we consider that $\xi^{\prime}$ and $\xi$ have the same variance $\sigma_2^2$.

Using MLE, the camera fingerprint can now be estimated from $g$ SDA-frames $I_1^{SDA}, I_2^{SDA}, \dots, I_g^{SDA}$ as
\begin{align} \nonumber
 \hat{K^{SDA}} = \frac{\sum_{i=1}^{g} W_i^{SDA} . I_i^{SDA}}{\sum_{i=1}^{g} \big(I_i^{SDA}\big)^2}.
\end{align}

Using Cramer-Rao Lower Bound, the variance of the estimated fingerprint $\hat{K^{SDA}}$ becomes
\begin{align} \label{eq:OurVar}
 Var (\hat{K^{SDA}}) \ge \frac{ \frac{\sigma_1^2}{d} + \sigma_2^2}{\sum_{i=1}^g \big(I_i^{SDA}\big)^2}.
\end{align}

In an ideal case, we want that the averaging operation does not degrade the quality of the estimated PRNU from the SDA-frames. In other words, we want that $Var (\hat{K^{SDA}})$ is approximately equal to the variance from the conventional method $Var (\hat{K})$. That is, 
in other words, using the results from (\ref{eq:conv1}) and (\ref{eq:OurVar}), it is desired that
$$
\frac{ \frac{\sigma_1^2}{d} + \sigma_2^2}{\sum_{i=1}^g \big(I_i^{SDA}\big)^2} \approx \frac{\sigma_1^2 + \sigma_2^2}{\sum_{i=1}^n I_i^2}.
$$
By simplifying the above equation, we get
$$
\frac{ \frac{\sigma_1^2}{d} + \sigma_2^2} {\sigma_1^2 + \sigma_2^2} \approx \frac {\sum_{i=1}^g \big(I_i^{SDA}\big)^2} {\sum_{i=1}^n I_i^2}.
$$

Suppose 
$$\frac {\sum_{i=1}^g (I_i^{SDA})^2} {\sum_{i=1}^n I_i^2} = \frac{g}{n} \times k$$ 
where 
$$k = \frac {(\sum_{i=1}^g (I_i^{SDA})^2)/g} {(\sum_{i=1}^n I_i^2)/n}.$$ Note that the value of $k$ is a temporary variable that is less than or equal to $1$ as the numerator $\sum_{i=1}^g (I_i^{SDA})^2)/g$ is less than equal to the denominator $\sum_{i=1}^n I_i^2)/n$. 
Putting these values in the above equation, we get
$$
\frac{g}{n} \times k \approx \frac{ \frac{\sigma_1^2}{d} + \sigma_2^2} {\sigma_1^2 + \sigma_2^2}.
$$
Putting $g=\frac{m}{d}$ in the above equation, we get
$$
\frac{m \times k}{d \times n} \approx \frac{\sigma_1^2 + d \times \sigma_2^2 }{d \times (\sigma_1^2 + \sigma_2^2)}.
$$
or,
\begin{align} \label{eq:finalourapproach}
 m \approx \frac{n}{k} \times \frac{\sigma_1^2 + d \times \sigma_2^2 }{\sigma_1^2 + \sigma_2^2}
\end{align}

We then discard the temporary variable, $k$, from the equation. Since $0 < k \leq 1$, the final equation becomes 
\begin{align} \label{eq:finalourapproach2}
 m \leq n \times \Big( \frac{\sigma_1^2 + d \times \sigma_2^2 }{\sigma_1^2 + \sigma_2^2} \Big)
\end{align}

From (\ref{eq:finalourapproach2}), we can derive the following concluding remarks:
\begin{itemize}
    \item Since $d \ge 1$, the right hand side of the equation is at least $1$. Therefore, the number of images required in the proposed SDA method (i.e., $m$) will be more than or equal to the number of images required in the conventional method (i.e., $n$).
    \item For smooth images $\sigma_2^2$ is close to zero. So, the impact of SDA-depth, $d$, will be negligible for such images. Therefore, SDA and conventional approaches will have similar performance. However the SDA technique will be $d$ times faster in the best case.
    \item For textured images, when the number of for both techniques is equal (i.e., $m = n$), because $\sigma_2$ is greater than zero, conventional approach is expected to outperform SDA approach. 
    \item Since $\sigma_2^2$ is greater than zero for textured images, the ratio of images for SDA- divided by conventional approach, $\frac{m}{n}$, will increase as the SDA-depth, $d$, increases. Therefore, SDA approach will require more images to achieve same performance for textured images.
\end{itemize}

Notice that it is hard to characterize the relationship of $\sigma_1$ and $\sigma_2$, also $\sigma_1$ depends on various factors such as shot noise, exposure time, temperature, illumination, image content and so on. Therefore, we are not focusing on their relationship in this research. In the following section, we experimentally validate the observations listed above.

%% file: sections/5a_exp_validation.tex
\section{Validation of analysis}
\label{sec:exp:valid}

In this section, we experimentally verify the main conclusions arrived at by the analysis performed in the previous section. In our experiments we use both flatfield and textured images from the VISION dataset~\cite{shullani2017vision}. The implementations were done using Matlab 2016a on Windows 7 PC with 32 GB memory and Intel Xeon(R) E5-2687W v2 @3.40GHz CPU. The wavelet denoising algorithm \cite{mihcak1999low} was used to obtain fingerprint and PRNU noise. PCE and NCC methods were used for comparison. A preset threshold of 60 \cite{goljan2009large} was used for PCE values. Values higher than this threshold were taken to conclude that the two media objects originated from the same camera.

\subsection{Studying the effect of smoothness}
To verify the observations of the analysis related to smoothness of the images used to compute a camera fingerprint, we randomly selected 50 flatfield images and $50$ textured images from each camera in the dataset. For each of these types, five experiments were conducted by using a random set of $5$, $10$, $20$, $30$, and $50$ images for computing the fingerprint. So for example, when we chose $30$ flatfield images, we created one fingerprint using the conventional approach by denoising each of the 30 images and then averaging the PRNU noise patterns to arrive at the fingerprint estimate. Then a fingerprint estimate using the SDA approach was computed by averaging the same $30$ images in the spatial domain first and then denoising this SDA-frame of depth $30$ to directly arrive at another fingerprint estimate. Therefore, a total of $20$ fingerprints were obtained for each camera ($2$ types of images; $2$ fingerprint extraction techniques; $5$ different cardinalities of image sets used for fingerprint computation). 

Each of these two fingerprints was correlated with the PRNU noise obtained from the rest of the images in the dataset taken with the same camera. This set consisted of both textured and flat-field images. 
To create an abundance of test cases, we divided each full resolution fingerprint into $500\times 500$ disjoint blocks and correlated them with the corresponding blocks in the test images to match the PRNU noise. As a result, a total of $244,127$ comparisons were made.


\begin{figure} [!ht]
\centering
{\includegraphics[width=\linewidth, trim={3.3cm 10.45cm 3.4cm 10.55cm},clip] {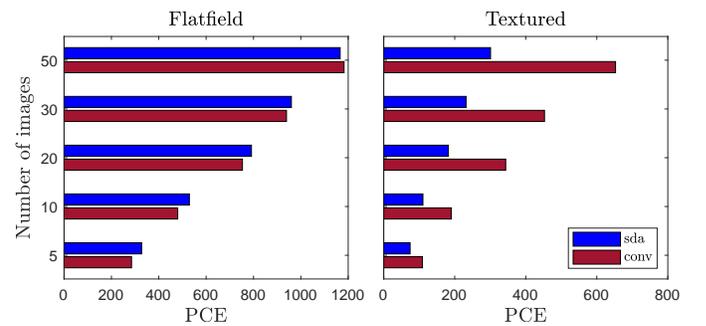}}
\caption{The effect of texture in terms of PCE}
\label{fig:exp:texture:pce}
\end{figure}

Fig.~\ref{fig:exp:texture:pce} shows how image content affects the PCE for fingerprints obtained from $5,10,20,30$ or $50$ flatfield and textured images. The figure shows that with flatfield images, despite the significantly lower number of denoising operations performed by the SDA approach, the results obtained are similar to the conventional approach. This observation holds regardless of the number of images averaged for fingerprint extraction. 
The performance of the SDA approach drops for textured images. However, this difference can be overcome by increasing the number of images used for SDA technique but still keeping the number of denoising operations lower than the conventional approach. We investigate this issue in the next subsection.

\begin{figure} [!ht]
\centering
{\includegraphics[width=\linewidth, trim={3.3cm 10.45cm 3.4cm 10.55cm},clip] {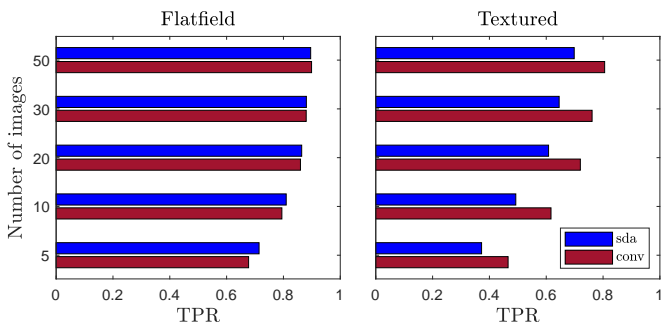}}
\caption{The effect of texture in terms of TPR}
\label{fig:exp:texture:tpr}
\end{figure}

If we consider the above results in terms of TPR, the SDA approach starts doing better as the PCE is thresholded around a set value (60 in our case) to arrive at the attribution result. So a drop in PCE does not necessarily result in a wrong decision. This improvement can be observed in Fig.~\ref{fig:exp:texture:tpr} which shows TPR for the same experiments when the threshold is set to $60$ as proposed in~\cite{goljan2009large}. 
The other implications of these figures are already well-known in the field (i.e., flatfield images are better than textured and as the number of images increase quality of fingerprint also increases which results in a higher PCE and TPR.)

Table~\ref{table:time:flat} shows the average time it takes to extract a fingerprint estimate by the two methods in the above experiment. Notice that in both cases the same number of images, $m$, are read from the disk but for the SDA technique only one denoising operation is needed whereas for conventional way, $m$ denoising operations are done. This implies that as the training images increase, the speedup also increases. 
A speedup of $13.5$ times can be achieved by averaging $50$ images before denoising. 

\begin{table}[!ht]
\centering
\caption{Average time to extract fingerprints with proposed and conventional methods (in sec)}
\label{table:time:flat}
\begin{tabular}{|c|c|c|c|c|c|}
\hline
 & $5$ & $10$ & $20$ & $30$ & $50$ \\ \hline
SDA & 4.97 & 5.99 & 8.22 & 10.35 & 14.49 \\ \hline
Conventional & 21.57 & 40.81 & 79.96 & 118.79 & 196.59 \\ \hline
Speedup & 4.34 & 6.81 & 9.73 & 11.48 & 13.57 \\ \hline
\end{tabular}
\end{table}

\subsection{Fingerprint equivalence for textured images}
For textured images, our analysis indicated that more images are needed by the SDA method and hence a corresponding reduction in the speedup obtained would occur. In this experiment, our goal is to investigate the relationship between the number of images required by SDA compared to the number needed by the conventional approach to yield similar performance for textured images while still retaining a speed-up in fingerprint computation. This experiment was again performed using images from the VISION dataset~\cite{shullani2017vision}. 

We created a training set from $50$ textured images for each camera in the VISION dataset. $19$ fingerprints were created using $2, 3, \dots 20$ images using the conventional approach. We also created $49$ fingerprints using SDA method using $2, 3, \dots 50$ images.
As done in the previous experiment, each fingerprint was partitioned into disjoint $500\times 500$ blocks and correlations were computed with the corresponding blocks of the test PRNU noise pattern. 


\begin{figure} [!ht]
\centering
{\includegraphics[width=\linewidth, trim={3cm 9cm 2.9cm 9cm},clip] {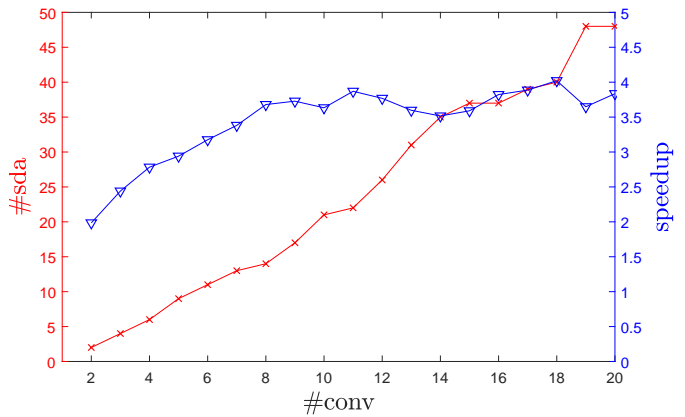}}
\caption{Fingerprint equivalence for SDA and conventional approaches. x-axis indicates number of images for conventional. The left of y-axis (red) is the number of images required for SDA and the right one (blue) is the speedup gained in this case.}
\label{figure:exp:fp_equi}
\end{figure}

Figure~\ref{figure:exp:fp_equi} shows the number of images required by the SDA approach to achieve at least the same TPR as the conventional approach. Moreover, it shows the speedup gained in these cases. 
For example, when fingerprint is created from $20$ textured images using conventional way, the same TPR can be achieved using $48$ images in SDA approach. In this way, the fingerprint extraction is approx. $3.85$ times faster for SDA approach. The figure shows that using $2-3$ times more images for SDA method, up to $4$ times speedup can be achieved with no loss in TPR when the images are textured.

\subsection{Effect of SDA-depth on image fingerprint}
In Section~\ref{sec:approach}, we have shown that as the SDA-depth increases, when the number of images for fingerprint extraction is constant, the TPR is expected to drop. To verify this remark, we used $50$ textured images for fingerprint extraction. We didn't include any flatfield image in this set as flatfield images results in a negligible difference in performance between SDA and conventional fingerprints. 

We then created fingerprints using $50$ textured images from each camera in the VISION dataset. We set SDA-depth to $1, 2, 5, 10, 25$ and $50$. Therefore, we created $50, 25, 10, 5, 2,$ and $1$ SDA-frames, respectively. The SDA-frames were denoised and then averaged to arrive at the final fingerprint estimate. For each fingerprint estimate computed, the rest of the images were used as test images. We correlated each fingerprint with the PRNU noise extracted from the test images in a block-wise manner as done in previous experiments. Notice that $SDA-1$ is the same as conventional approach. 

\begin{table}[!ht]
\centering
\begin{tabular}{|c|c|c|c|c|c|c|}
\hline
 & SDA-$1$ & SDA-$2$ & SDA-$5$ & SDA-$10$ & SDA-$25$ & SDA-$50$ \\ \hline
PCE & 652.8 & 514.6 & 390.0 & 332.2 & 285.0 & 252.4 \\ \hline
TPR & 0.80 & 0.78 & 0.75 & 0.72 & 0.69 & 0.67 \\ \hline
\end{tabular}
\caption{SDA-depth vs TPR and PCE, change with figure}
\label{table:sda_depth_perf}
\end{table}

Table~\ref{table:sda_depth_perf} shows that as the SDA-depth increases, the average PCE decreases. 
For textured images, the more images we combine to create an SDA-frame, the lower the PCE and TPR values that will result. This supports the third observation of the analysis in Section 3. 

This section has provided a validation of Section~\ref{sec:approach} by experimentally supporting all three observations derived from the analysis. Namely, when images are not textured, hence resulting in low post-filtering noise, both the SDA and conventional fingerprints from the same images perform similarly which can lead to $13.5$ times speedup. On the other hand, textures images and larger SDA-depth result in requiring higher number of images to achieve the same performance as conventional approach. Yet, a speedup by a factor of $4$ can still be achieved in most cases. 

In the next section, we apply the proposed approach to practical problems, and show that SDA fingerprints can perform with a significantly higher accuracy or result in significant speedup compared to state-of-the-art fingerprint extraction techniques.

%% file: sections/5b_exp_video.tex
\section{Application to computing video fingerprints}
\label{sec:exp:application}

In this section, we investigate a more practical use case of the proposed SDA technique which is its usage for extracting FE from videos. As Section~\ref{sec:background} explains, two of the most common ways to extract a fingerprint from a video are using only I-frames or using all frames (or the first $n$ frames). While the former results in low performance, the latter can be impractical in many real life applications due to very high computational needs. For example, fingerprints from $50$ $1-$minute videos (i.e., approximately $1800$ frame per video) using a single-thread may take up to a day to compute. In this section, we provide experimental results that demonstrate how using the SDA approach can provide significant improvements in the time needed for computing fingerprint estimates from video, while retaining the same performance obtained using a significantly larger number of denoising operations using conventional approaches. 

In each experiment below, three different types of fingerprints (i.e., I-frames only, SDA-frames and ALL-frames) were obtained from each video.  For the sake of simplicity, we refer to them as \textit{I-FE} (i.e., \textit{Fingerprint Estimate}), \textit{SDA-FE}, and \textit{ALL-FE}, respectively. Moreover, in some cases, we add an indication of the SDA-depth when we need to highlight it. For example, SDA-50-FE indicates that the video frames were divided into groups of $50$ and each group averaged to create an SDA-frame.

In the first experiment, we examine source matching for videos. That is given two videos, can we determine if they are from the same camera. Next we investigate a more difficult case that involves mixed media. In this subsection, we also analyze an important question related to mixed media: ``What is a good balance of SDA-depth which optimizes speed and performance?". 
In the next two subsections, we examine the performance achieved with video and images obtained from social media such as Facebook and YouTube. Finally, we show how the proposed technique can be used for source attribution with moderate length stabilized videos (i.e., up to $4$ minutes) from which obtaining a ``reliable" FE might take couple of hours each using all frames. 

Two datasets were used in all the experiments, the NYUAD-MMD, and VISION datasets. The NYUAD-MMD dataset contains images and videos of different resolutions and aspect ratios from $78$ cameras from different models and brands. This makes it a challenging dataset for mixed media attribution. Moreover, it contains stabilized videos longer than $4$ minutes from $5$ cameras. Hence, we used this dataset for experiments using mixed media and stabilized video.
The videos in the dataset are typically around $40$ seconds( i.e., each video is approximately $1200$ frames) and images are pristine (i.e., no out-camera operations). 
The VISION dataset contains different high quality videos and images from social media such as Facebook and YouTube. Hence, we used this dataset in experiments involving social media.

\subsection{Matching Two Non-Stabilized Videos}
\label{exp:vision:two_vid}
In the first experiment, we examine source matching for videos using FE computed from the three different approaches that have been presented.  Our goal was to estimate the length of videos and the resulting computation time needed to achieve greater than $99\%$ TPR for I-FEs, SDA-FEs and ALL-FEs. This way, a clear comparison of the the three approaches could be made.

FE from the non-stabilized videos of the same resolution from the VISION dataset were first created. FE were extracted from the first $5, 10, \dots40$ seconds of each video using the two techniques mentioned in Section~\ref{sec:background} and the proposed method. On average, each video had approximately one I-frame per second. We selected an SDA-depth of $30$ resulting in an SDA-frame from each second of video. 

\begin{figure} [!ht]
\centering
{\includegraphics[width=\linewidth, trim={2.9cm 9.5cm 2.8cm 9.5cm},clip] {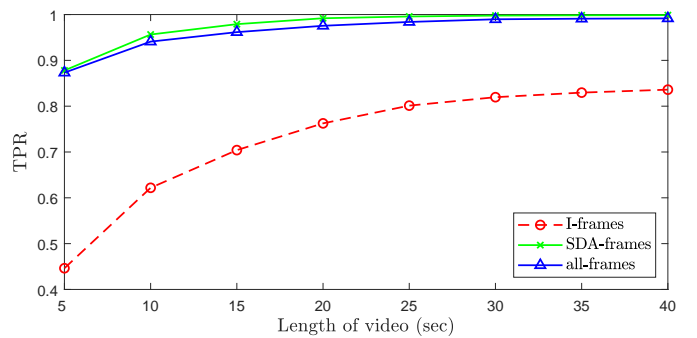}}
\caption{TPR for different lengths of video using I-FEs, SDA-FEs, and ALL-FEs}
\label{fig:exp:two_vid}
\end{figure}

Figure~\ref{fig:exp:two_vid} shows TPR using I-FE, SDA-FE, and ALL-FE as the length of the videos increases. As seen, SDA-FEs outperforms ALL-FEs in this setting for all video lengths. The difference varies between $0.5$ (for $5$ sec videos) and $1.7\%$(for $15$ sec videos). 
Both FE achieve significantly higher TPR than I-FEs. For example, for $10$ seconds video, SDA-FEs and ALL-FEs result in $94.1\%$ and $95.6\%$ TPR, respectively, whereas I-FEs can only reach $62.2\%$ TPR. 

The highest TPR achieved using I-FEs was $83.7\%$ (i.e., for $40$ second videos) which is still lower than the TPR of SDA-FEs and ALL-FEs when they were computed from only $5$-second videos (i.e., more than $87\%$). This is because SDA-FE and ALL-FEs use all the $150$ frames in a 5 second video (i.e., I-, B- or P-frames) whereas the I-FEs use only $40$ I-frames on average and ``waste" the rest of the frames. Hence, for this setting, I-FEs fail to reach to a comparable accuracy as the other two methods. 

\begin{table}[!ht]
\centering
\caption{Time for video fingerprint extraction in second}
\vspace{-0.3em}
\begin{tabular}{|c|c|c|c|}
\hline
type      & averaging & I/O + denoising & total  \\ \hline
I-FE    & 0     & 50       & 50    \\ \hline
SDA-FE   & 12     & 50       & 62    \\ \hline
ALL-FE   & 0     & 1407      & 1407   \\ \hline
\end{tabular}
\label{tbl:vF_extraction}
\vspace{-1em}
\end{table}

We then estimated the time required for extraction of each FE from a $40$ second Full HD video captured @30 FPS.
Table~\ref{tbl:vF_extraction} compares the average times for them. It takes $50$, $62$, and $1407$ seconds for an I-FE, SDA-FE and ALL-FE, respectively. However, these times are when each one is obtained from $40$ second videos. When we evaluate the required time to achieve $83\%$ TPR, we need less than $5$ seconds of video for SDA-FEs and ALL-FEs whereas I-FEs require $40$ seconds of video. This suggest that the required time for SDA-FEs and ALL-FEs are less than $8$ and $176$ seconds, respectively. Hence, SDA technique is at least $6$ times faster than I-FEs and requires $8$ times shorter videos, yet still achieves a higher TPR. Moreover, it performs up to $1.7\%$ higher than ALL-FEs in terms of TPR and speeds up approximately $22.5$ times in this setting. Moreover, while SDA-FEs can achieve $99\%$ TPR with $20$ seconds videos, the same can be achieved with $30$ seconds for ALL-FEs. Therefore, close to $34$ times speedup can be achieve in this case when SDA-depth is set to $30$.

Notice that these results involve videos that did not undergo any processing such as scaling, compression in social media and so on. Also, all videos were taken with high luminance in the VISION dataset. Therefore, it is possible to have lower performance with more difficult datasets such as when videos are dark or processed. However, our intention here was to demonstrate the effectiveness of SDA approach first for the simplest of cases. We examine more challenging situations in further experiments below.

\subsection{Mixed Media Attribution}

As we have seen in the previous subsection, using I-FEs causes a significant drop in TPR whereas $20-30$ seconds of video is enough to achieve more than $99\%$ TPR for both SDA-FEs or ALL-FEs. In this subsection, we investigate a more challenging scenario where a video FE needs to be matched with a single query image. In \cite{taspinar2019source}, source attribution with mixed-media was investigated using the NYUAD-MMD dataset which is a very challenging dataset containing images and videos of various resolutions from $78$ of cameras. Here, we performed ``Train on videos and test on images" experiment for I-FEs, SDA-FEs, and ALL-FEs. That is a camera FE was computed from the video and the query image was cropped and resized and its PRNU matched with the FE. The resizing and cropping parameters to perform the matching were obtained from the ``Train on images, test on videos" experiment done in ~\cite{taspinar2019source}). 

The videos in this dataset were typically around $40$ seconds long; each having approximately $1200$ frames. The dataset contains a total of $301$ non-stabilized videos and $6892$ images from those cameras.
Each video FE was correlated with the PRNU noise of all the test images from the same camera to estimate ``true cases" which ended up with $23571$ correlations. Then, each video FE from $i^{th}$ camera was compared with the PRNU noise of images from $(i+1)^{th}$ camera for resizing and cropping parameters that maximizes the PCE for the image FE (i.e., the FE obtained from all images of the camera using conventional approach). This way, we estimated the ``false cases" resulted in $17755$ correlations.

\begin{figure} [!ht]
\centering
{\includegraphics[width=\linewidth, trim={2.7cm 10.2cm 2.6cm 9.5cm}, clip]{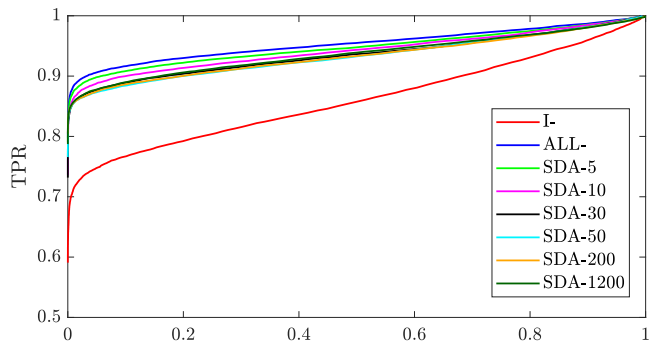}}
\caption{The ROC curves for varying SDA-depths}
\label{figure:experiment:num_image_sda}
\end{figure}

In the previous experiment we had used a fixed SDA-depth, $d$, of $30$. In this experiment we used different SDA-depths to investigate its impact on performance and speed. 
Given a video of $m$ frames (in our case approximately $1200$ frames), we divided the frames into groups of $d = 1, 5, 10, 30, 50, 200, 1200$. Therefore, the number of SDA-frames, $g$, became $1200, 240, 120, 40, 24, 6, 1$ respectively. When $g = 1$, the technique becomes the same as using all frames whereas when $p = 1200$, only a single SDA-frame is created by averaging all $1200$ frames. After obtaining the PCE of the ``true" and ``false" cases, we created an ROC curve for each video FE type/depth. Figure~\ref{figure:experiment:num_image_sda} shows the ROC curves for each of the SDA-FEs of different depths, as well as I-FE and ALL-FE. The results show that ALL-FE results in the highest performance, whereas I-FE perform significantly poorer compared to others. The proposed SDA method performs close to ALL-FE method for all depths. 

Table~\ref{tbl:sda_depth} shows more detailed results. $|\overline{PCE}|$ stands for the average of the PCE ratios with respect to I-FEs. For example, when an ALL-FE from $i^{th}$ video is correlated with the noise of $j^{th}$ image, its PCE is on average $3.2\%$ times higher compared to the $I-FE$ obtained from the same video. The reason we used such a normalization instead of average PCE is that outliers have a big impact on average PCE. Moreover, the table shows the TPR for the PCE threshold of $60$, average time to extract a FE, and the speedup compared to ALL-FEs. As seen, the results indicate that the TPR of SDA method are very close to ALL-FE. However, a speedup of up to $52$ times can be achieved using the SDA method.

\begin{table}[!ht]
\centering
\renewcommand*{\arraystretch}{1.15}
\caption{Detailed information for mixed media attribution}
\label{tbl:sda_depth}
\begin{tabular}{|c|c|c|c|c|c|c|c|c|}
\hline
      	  &I-  & ALL- & 5  & 10  & 30  & 50  & 200 & 1200 \\ \hline
$|\overline{PCE}|$ & 1.0 & \textbf{3.2}  & 3.1 & 2.9 & 2.6 & 2.6 & 2.5 & 2.4 \\ \hline
TPR($\%$)	   &64.0 & \textbf{83.1} 	& 82.3	& 81.3 & 80.0 & 79.8 & 80.1 & 79.8 \\\hline
time(s)     & 50  & 1407	& 276	& 142 & 62  & 48  &	32	& \textbf{27}  \\ \hline
speedup     &28.1 & 1.0 & 5.1 & 9.9 & 22.7 & 29.3 & 44.0 & \textbf{52.1}\\ \hline
\end{tabular}
\end{table}

Similar to the previous experiment using I-FEs have significantly lower accuracy (at least $16\%$ lower TPR). Moreover, when SDA-depth $\geq 30$, SDA-FEs are faster to extract as compared to I-FEs. Notice that when ALL-FEs are used, it takes approximately five days to extract all the FEs from the $301$ videos in the NYUAD-MMD dataset using a single-threaded implementation. This type of performance will clearly impractical  for many applications.

\subsection{Train and test on YouTube videos}

This experiment explores the performance achieved when two video FEs from YouTube are correlated. Although this experiment is essentially the same as the Section~\ref{exp:vision:two_vid}, it is relevant in practice as high compression is involved. Note that a key motivation of the SDA approach is that when high compression is used, a large number of frames are needed for computing a reliable FE. We created FE from all non-stabilized YouTube videos in VISION dataset (i.e., the ones labeled flatYT, indoorYT, and outdoorYT) using only I-frames, SDA-$50$, SDA-$100$, SDA-$200$, and ALL-frames. Here, we used the first $10, 20, \dots 60$ seconds of the YouTube videos to extract FEs. Each $60$ second video had approximately $1800$ frames that were used for SDA- or ALL-FEs, whereas they contained $31.3$ I-frames on average. After fingerprint extraction, we correlated each video FE with others of the same type and same length taken by the same camera. For example, an I-FE from $20$ seconds of video is correlated with all I-FEs obtained from the rest of the $20$ seconds videos from the same camera.
The same was done for SDA- and ALL-FEs. This way, a total of $3124$ correlations were done for each type.

\begin{figure} [!ht]
\centering
{\includegraphics[width=\linewidth, trim={2.9cm 9.7cm 2.8cm 10cm},clip] {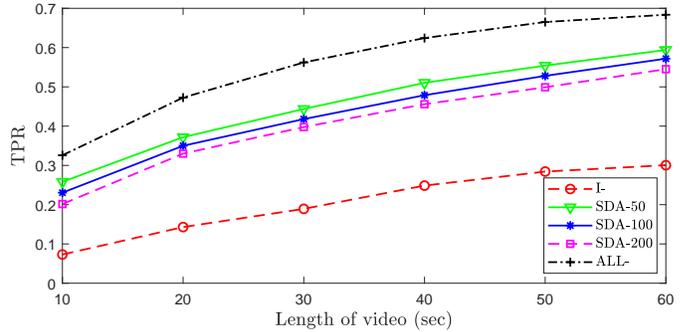}}
\caption{The effect of FE type and video length on TPR for YouTube videos}
\label{fig:exp:youtube}
\end{figure}

Figure~\ref{fig:exp:youtube} shows the TPR for varying lengths of video for each FE type. The figure shows that I-FEs perform very poorly for all cases and any FE type created from video of more than $20$ seconds outperforms I-FEs. While ALL-FEs perform better than SDA-FEs for the same-length videos, this difference can be overcome by increasing the video length but still using much fewer denoising operations. For example, SDA$-50$ obtained from $50$ second videos or SDA$-100$ from $60$ seconds videos, perform approximately the same as ALL-FEs obtained from $30$ seconds (within $\mypm1\%$ TPR range). Hence, instead of using $900$ frames for ALL-FEs, using $1800$ frames for SDA$-100$ can result in significant speedup with no loss in TPR. While an ALL-FE from $900$ frame of a Full HD video takes $1045$ seconds to compute, and SDA$-100$ FE from $1800$ frames, which only does $18$ denoising instead of $900$, takes $56$ seconds to compute. Therefore, a speedup of close to $19$ times can be achieved with SDA$-100$ with $1\%$ increase in TPR. Notice that, because most videos are around $60$ seconds in the VISION dataset, it limits the maximum length we could use in our experiments.

\subsection{Train on Facebook images, test on YouTube videos}
From the previous experiments, we know that the SDA method can help achieve a significant speedup for both videos and images with a small loss in performance which can be overcome by increasing the number of still images used for fingerprint extraction if available. In this experiment, our goal was to show that the proposed method can be successfully applied to other social media. Specifically, in this subsection, we extract FEs from Facebook images and match them with the FE of YouTube videos. We call this the ``Train on Facebook images, test on YouTube videos" experiment. The importance of this experiment is both media sharing services contain billions of visual media and computing ALL-FEs from these collections can have very high time complexity. Therefore, faster fingerprint extraction methods (along with search techniques) that speeds up attribution are badly neededl

In this experiment, for the cameras in the VISION dataset that had non-stabilized videos, we created a FE from $100$ Facebook images (i.e., the ones labeled FBH) using conventional fingerprint computation method. We then used the FEs from non-stabilized YouTube videos (those created in the previous experiment). We again used I-frames, SDA-$50$, SDA-$200$, SDA-$600$, and ALL-frames that were computed from the first $60$ seconds of YouTube videos. We then correlated the image FE of a camera with the FE of each video of each type using the efficient search proposed in ~\cite{taspinar2019source} and a total of $343$ pairs were compared for each FE type. Table~\ref{tbl:exp:ns:fb_yt} shows the TPR of these correlations. Similar to ``Train on videos, test on images" experiment, these results show that for FEs obtained from Facebook images matches with $81.34\%$ TPR with the YouTube videos for SDA-$50$ which is higher than both ALL-FEs and I-FEs. On the other hand, FEs from I-frames yield approximately $30\%$ lower TPR. These results show that SDA approach is a good replacement over using I-FEs or ALL-FEs for this scenario.

\begin{table}[!ht]
\centering
\caption{TPR of different FE types when a FE from Facebook images and another from YouTube videos are correlated}
\label{tbl:exp:ns:fb_yt}
\begin{tabular}{|c|c|c|c|c|c|}
\hline
  & I-FE  & SDA-$50$ & SDA-$200$ & SDA-$600$ & ALL-FE  \\ \hline
TPR & 51.60 & 81.4   & 79.88   & 78.13   & 79.59  \\ \hline
\end{tabular}
\end{table}

\subsection{Matching two stabilized videos}
A recent work \cite{sri2018stabilization} has shown that a FE obtained from a long stabilized video can successfully be matched with other videos from the same camera. However, thousands of frames must be denoised. This may not be practical in many circumstances. A potential alternative for this problem is the use of SDA method which may lead to a significant speed up. To evaluate this, we captured stabilized videos from $5$ cameras. A total of $37$ videos were captured which added up to $260$ minutes. 

We extracted FEs from the frames of $20, 40, \dots 240$ second video lengths using conventional (I-frame and ALL-Frame) method as well as SDA method for SDA-depths of $30, 50,$ and $200$. These depths were deemed to be reasonable choices from previous experiments. As shown in \cite{taspinar2016source, iuliani2017hybrid, mandelli2019facing}, the first frame of the videos are typically not geometrically transformed. Since we divide video into pieces, some video pieces do not have an untransformed frame. So, we discarded the first frame of each video to avoid inconsistencies. 
We correlated each FE with the other FEs of different videos from the same camera that are created using the same number of frames. For example, $SDA-30-$FEs of $20$ second videos are correlated with the same type FEs from the same camera. 

\begin{figure} [!ht]
\centering
{\includegraphics[trim={1.2cm 8.4cm 1.4cm 8.7cm}, clip, keepaspectratio=true,width=\linewidth]{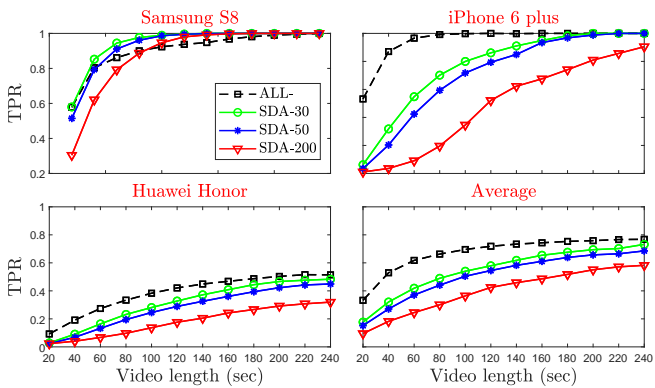}}
\caption{TPR for stabilized videos for varying SDA-depths}
\label{figure:exp:stab_res}
\end{figure}

Figure~\ref{figure:exp:stab_res} shows the TPR for three cameras (i.e., Huawei Honor, Samsung S8, and iPhone 6plus) and the total average of all the five cameras.

The results show that as videos get longer, ALL-FEs and SDA-FEs achieve higher TPR. Moreover, the effect of increased SDA-depth is more significant for this case in comparison to non-stabilized videos. 
While for some cameras ALL-FEs and SDA-FEs perform similarly (e.g., Huawei and Samsung cameras), for others (e.g., iPhone cameras) there is a significant difference between the two. For example, for Samsung S8 $SDA-200$-FE from $120$ seconds video, perform similarly as $180$ seconds ALL-FE. Therefore, for this particular case, $SDA-200$ can speedup $66$ times \big(i.e. $\frac{180}{120}\times \frac{1407}{32}$\big)(see Table~\ref{tbl:sda_depth} for times). On the other hand for iPhone 6 plus, ALL-FEs from $60$ seconds video and $160$ seconds $SDA-50$ have similar TPR. Therefore, $11$ times \big(i.e., $\frac{60}{160}\times \frac{1407}{48}$\big) speedup can be achieved in this case. Hence, a speedup between these numbers (i.e. $11$ and $60$) can be achieved without any loss in TPR if a long video is available.

Overall, this section shows that the proposed SDA-FEs outperforms the commonly used I-frame-only technique in all the cases for videos. These include mixed media, stabilized videos, and social media. On the other hand, the SDA-FEs achieves comparable results as ALL-FEs with up to $52$ times speedup in these experiments. We also show the impact of SDA-depth on the performance that can be achieved in various cases.

%% file: sections/7_conclusion.tex
\section{Conclusion and Future Work}
\label{sec:conclusion}

This paper has investigated camera fingerprint extraction using Spatial Domain Averaged frames, which are the arithmetic mean of multiple still images. By adding one extra step of averaging before denoising, a significant speedup can be achieved for fingerprint extraction. We show that this technique can successfully be used for images, non-stabilized videos as well as stabilized video to speedup fingerprint extraction process. The proposed method is especially useful when the number of denoising operations needed can be very high. For example, when dealing with non-stabilized or highly compressed stabilized videos or images from social media. 

It is often considered that for video source attribution, using only I-frames for fingerprint extraction (I-FEs) is ``enough" to achieve high performance. However, in this research, we have shown that I-FEs performs poorly compared to ALL-FEs in all cases. On the other hand, using ALL-FEs is impractical due to the large computation time needed for practical scenarios where thousands of videos can be available. The proposed SDA approach comes into play here to resolve the problem of I-FEs (i.e., accuracy) and ALL-FEs (i.e., speed). Both SDA- and ALL-FEs perform similarly in most cases. When the SDA method performs worse, this can be overcome by using more of the available frames if any. 
 
The proposed technique can be used for other source attribution related problems where many denoising operations are needed. For instance, this method can be applied when many ``partially misaligned" still images and a suspect camera are available. For example, a seam carved video contains many partially misaligned frames with its source camera. In such a scenario, instead of denoising all frames of the video, the SDA technique can be used as a way to speed up this process. 
Moreover, determining whether a video is stabilized or not is another issue which requires a number of denoising operations. As an alternative to using only I-frames, the proposed SDA technique could successfully work with only $2$ denoising operations. 

Another avenue for future research is to create an SDA-FE in a weighted manner such that performance achieve with SDA method can be increased. Two of the potential ways to achieve this are weighting I-, P- and B- frames differently, and weighting the frames in a block-by-block manner. For example, it has been shown that flatfield images perform better with SDA method compared to textured ones. Using this idea, one may weight textured regions differently from smooth regions.